\documentclass[aps,pra, superscriptaddress, reprint, longbibliography]{revtex4-1}

\usepackage{lipsum}
\usepackage{graphicx}
\usepackage{bm}
\usepackage{fancyref}
\usepackage{soul}
\usepackage[colorlinks,
			linkcolor=blue,
			urlcolor=blue,
			citecolor=blue]{hyperref}
\usepackage[T1]{fontenc}       
\usepackage[version=3]{mhchem}
\usepackage{xcolor}
\usepackage{enumerate}
\usepackage{titlesec}
\titlespacing*{\section}{0pt}{1.1\baselineskip}{\baselineskip}

\begin{document}
		
	\title[]{Reversible oxygen vacancies doping in $\mathbf{(La_{0.7},Sr_{0.3})MnO_3}$ microbridges\\by combined self-heating and electromigration}
	
	\author{Nicola \surname{Manca}}
	\thanks{Present address: Kavli Institute of Nanoscience, Delft University of Technology, P.O. Box 5046, 2600 GA Delft, The Netherlands}
	\affiliation{\mbox{Physics Department, University of Genova, Via Dodecaneso 33, Genova, Italy}}
	\affiliation{\mbox{CNR-SPIN, Corso Perrone 24, Genova, Italy}}	
	
	\author{Luca \surname{Pellegrino}}
	\email{luca.pellegrino@spin.cnr.it}
	\affiliation{\mbox{CNR-SPIN, Corso Perrone 24, Genova, Italy}}
		
	\author{Daniele \surname{Marr\'e}}
	\affiliation{\mbox{Physics Department, University of Genova, Via Dodecaneso 33, Genova, Italy}}
	
	\begin{abstract}
		Combination of electric fields and Joule self-heating is used to change the oxygen stoichiometry and promote oxygen vacancy drift in a free-standing $\mathrm{(La,Sr)MnO_3}$ thin film microbridge placed in controlled atmosphere.  By controlling the local oxygen vacancies concentration, we can reversibly switch our LSMO-based microbridges from metallic to insulating behavior on timescales lower than 1\,s and with small applied voltages (<5\,V). The strong temperature gradients given by the microbridge geometry strongly confine the motion of oxygen vacancies, limiting the modified region within the free-standing area. Multiple resistive states can be set by selected current pulses that determine different oxygen vacancies profiles within the  device . Qualitative analysis of device operation is also provided with the support of Finite Element Analysis.\\
		
		\noindent\textcopyright 2015 AIP Publishing LLC.\\
		Published online 18 May 2015 (\url{http://dx.doi.org/10.1063/1.4921342})
	\end{abstract}
	
	\maketitle
	
	The physics of oxygen vacancies in oxides has been matter of intensive investigation in the last decades \cite{Blanc1971} and still now	is the subject of euphoric research due to the increasing complexity of such systems showing an ever growing panorama of physical properties \cite{Morosan2012}. If, on one side, a great number of publications is focused on resistive switching phenomena at metal/oxide interfaces or in stacked multilayers for the development of memristive devices \cite{Waser2007, Waser2009, Jeong2012}, on the other hand, the use of oxygen vacancies for modulating the physical states of correlated oxides has been recently outlined \cite{Kalinin2013}. Also, the possibility of tuning the metal-insulator transition in vanadium dioxide \cite{Jeong2013} or in the two dimensional electron gas forming at the \ce{SrTiO3 /LaAlO3} interface \cite{Bark2012} by changing oxygen vacancies concentration has been reported. Many publications report on tuning of the physical properties of oxide films, for example, their resistance \textit{vs} temperature characteristics, as a function of oxygen stoichiometry, being the oxygen content modulated either by growing or by annealing the films in different oxygen pressure conditions \cite{Xie2014a}. Although these solutions offer the chance to conveniently and systematically explore the effects	of oxygen stoichiometry on oxide thin films, the exploitation of these capabilities into innovative electronic devices is limited by the impossibility of rapidly and reversibly changing	oxygen vacancies concentration over selected regions with low energy consumption. An alternative approach is to apply	an electrical current to move oxygen vacancies in stacked \cite{Fu2014,Yokota2013} or planar systems \cite{Balcells2013}. In the first case, contact interfaces also contribute to the overall resistance, making it difficult to isolate bare bulk effects, while both approaches require conventional thermal annealing of the whole sample to restore the initial conditions. To overcome these limitations, we propose the use of microhotplate technology, where a conducting oxide thin film works both as functional and heating element. The simple perovskite manganites $\mathrm{(A,B)MnO_3}$ family is a prototypical material showing competing electronic phases \cite{Tokura1996,Haghiri-Gosnet2003}. Manganites are characterized by complex temperature-dependent electronic phase diagrams determined by the A and B elements and their reciprocal ratio \cite{Imada1998,Gorkov2004}. Contrary to cation substitutions, doping by oxygen vacancies ($\mathrm{V_O}$) can be adopted to reversibly modify the physical properties of these oxide materials, such as magnetism, structural transitions, and electrical conductivity \cite{Mitchell1996,Renard1999,AktherHossain1999,Abdelmoula2000}. Also, the perovskite lattice structure is resilient to the incorporation of relatively large amount of defects such as oxygen vacancies \cite{Rekas2000,SaifulIslam2000,Rossetti2013}.
	
	\begin{figure*}[t]
		\begin{minipage}[c]{\textwidth}
			{
				\begin{minipage}[t]{0.7\textwidth}
					\mbox{}\\[-\baselineskip]
					\includegraphics[width=1.0\textwidth]{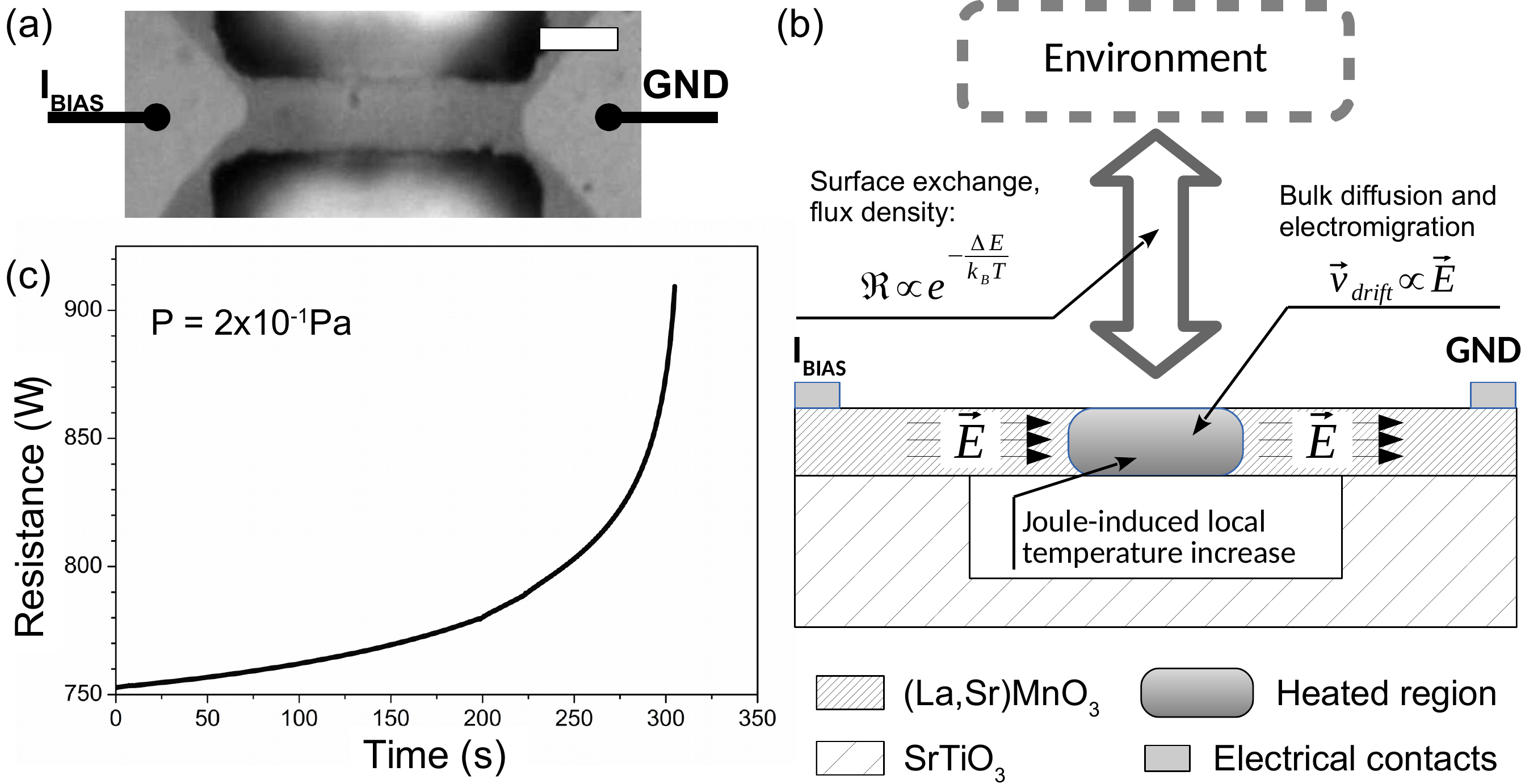}
				\end{minipage}\hfill
				\begin{minipage}[t]{0.25\textwidth}
					\mbox{}\\[-2\baselineskip]
					\caption{\label{fig:Fig1}
						Principle of device operation.
						(a) Picture of a typical device, darker regions are free-standing, while lighter areas are clamped to the \ce{SrTiO3} substrate. Electrical contacts are far from the shown pattern. White bar is 5\,$\mu$m.
						(b) Sketch of the basic mechanisms related to oxygen vacancies generation and transport inside the microbridge. The electric field is determined by the
						applied current bias at the electrical contacts.
						(c) Drift of the electrical resistance measured across the micro-bridge while biasing with 1.1\,mA in
						low vacuum conditions ($2\cdot10^{-1}$\,Pa).
					}
				\end{minipage}
			}
		\end{minipage}
	\end{figure*}
	
	In this work, we reversibly modify the physical properties of a prototypical correlated oxide, $\mathrm{(La_{0.7},Sr_{0.3})MnO_3}$ (LSMO), by changing its oxygen content through the application of an electrical current in a controlled environment. The use of free-standing LSMO microbridges is the key point of this work and brings peculiar features. For example, they can be heated
	to high temperature (up to melting) with small amount of Joule power in less than a millisecond in a well-defined localized region. This region can efficiently exchange oxygen with the environment due to the maximized surface-to-volume ratio and the effects of changes in the oxygen stoichiometry on the compound can be directly probed, ruling out interface	effects typical of metal-insulator-metal resistive switching devices \cite{Yang2008}. We exploit both the localized temperature increase provided by Joule self-heating and the applied electric fields as driving mechanisms for the oxygen vacancies.
	
	LSMO thin films were grown on \ce{STO(001)} substrates by pulsed laser deposition. Details on deposition conditions, microbridge fabrication, and their electro-thermal characterization have been already reported in a previous work, together	with the description of the experimental setup \cite{Ceriale2014}. In the present experiments, we used 5\,$\mu$m wide microbridges having thickness and length in the 50\,nm--100\,nm and 10\,$\mu$m--20\,$\mu$m range, respectively; a typical device is shown in Fig.\,\ref{fig:Fig1}(a). All the electrical measurements were performed in four probes configuration. The working principle of our experiment is sketched in Fig.\,\ref{fig:Fig1}(b). The evolution of $\mathrm{V_O}$ doping along the	microbridge is determined by two different mechanisms: the exchange of oxygen with the external environment at the surface and the diffusion/electromigration of the created oxygen vacancies within the bulk. The electrical current applied to the microbridge can be used both to read its electrical resistance value and to modify the oxygen vacancies doping condition. In the first case, we typically used 10\,$\mu$A bias, while in the latter higher currents are required (between 0.6\,mA and 1.2\,mA). We note that temperature of the whole sample (resulting in uniform heating of the microbridge) is changed by placing the substrate in contact with a thermostat. In the following, the reported temperatures are those of the thermostat, while the increase of microbridge temperature by Joule effect is not	directly measured, but estimated by Finite Element Analysis (FEA) \cite{Ceriale2014}. The applied electrical current determines an increase of temperature along the device with maximum value at its center, depending on the interplay between thermal couplings
	and Joule power density \cite{Ceriale2014}. The high temperatures activate oxygen ions bulk drift and surface exchange phenomena; thus, oxygen vacancies can be easily created by just lowering the	external \ce{O2} pressure or, alternatively, $\mathrm{V_O}$ can be recombined if \ce{O2} partial pressure is increased. The created (positively charged) $\mathrm{V_O}$ are displaced inside the lattice by the electric	field given by the current bias. $\mathrm{V_O}$ mobility is strongly temperature-dependent, thus in the center of the microbridge the electric field is much more effective in moving the newly formed vacancies. Under the effect of the electric field, the $\mathrm{V_O}$ approach to the clamping regions where they are frozen in place by the decreased temperature, given by the proximity of the substrate acting as heat sink.
	
\begin{figure}
		\includegraphics[width=0.9\linewidth]{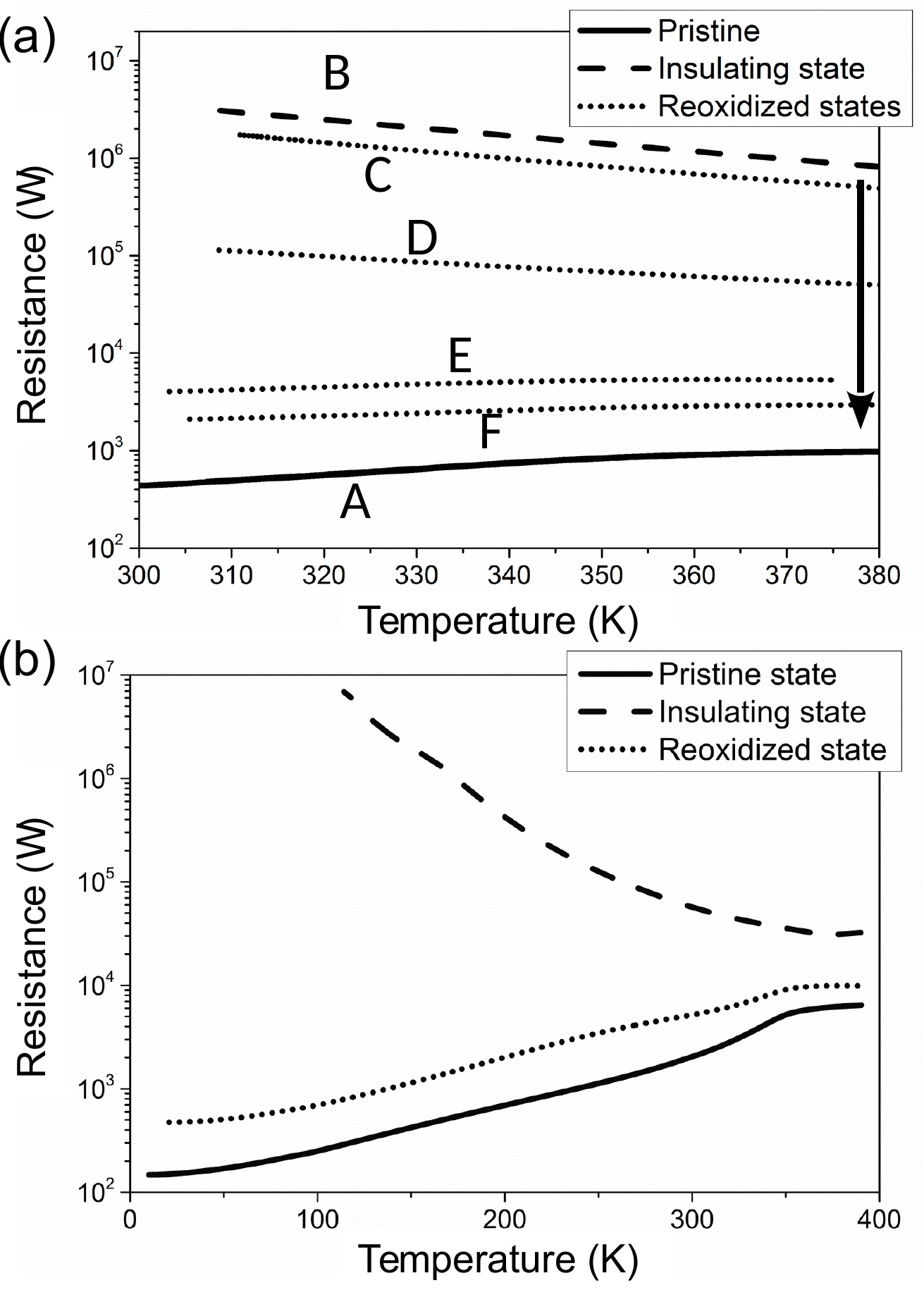}
		\caption{\label{fig:Fig2}
			Metal-insulator transition induced by oxygen vacancies doping in LSMO free-standing microbridges.
			(a) $R(T)$ plots of the reduced insulating state (B) is obtained by current biasing in low vacuum ($\mathrm{2\cdot10^{-1}}$\,Pa) at 300\,K from pristine sample (A), while intermediate states ((C)--(F)) by subsequent current pulses ($\mathrm{10\,\mu A-100\,\mu A}$) applied in air environment at 350\,K.
			(b) R(T) plot of a pristine, reduced and reoxidized LSMO microbridge showing the restored metallic behaviour in the full 4\,K--400\,K temperature range.
		}
	\end{figure}

	The progressive formation of $\mathrm{V_O}$ can be detected by measuring the time evolution of the electrical resistance of a
	LSMO microbridge under constant (1.1\,mA) current bias in low vacuum ($\mathrm{2\cdot10^{-1}}$\,Pa). We observe the divergent behavior typical of a thermal runaway process (Fig.\,\ref{fig:Fig1}(c)) which, in our case, is limited by the voltage compliance of the current	generator. LSMO resistivity is strongly affected by oxygen stoichiometry and an increase in $\mathrm{V_O}$ concentration significantly modifies the measured electrical properties, with a smooth transition from metallic to semiconducting behavior.
	Fig.\,\ref{fig:Fig2}(a) shows the R(T) curves in the 30\,$^\circ$C--100\,$^\circ$C range of the same microbridge after the application of different current pulses in controlled environment. At first, the pristine microbridge is biased (1.1\,mA) in low vacuum conditions with the voltage compliance matched to reach a high resistance state with insulating-type behavior (plot B of Fig.\,\ref{fig:Fig2}(a)). Successively, air environment is set again and the initial state is progressively recovered by current pulses of selected magnitude (plots C--F). In this experiment, we also kept sample at 80\,$^\circ$C to enhance $\mathrm{V_O}$ mobility and oxygen exchange rate at the device surface. We measured almost four orders-of-magnitude resistance change at 30\,$^\circ$C, but even higher values can be achieved by an accurate control of thermal runaway. The restored metallic state does not overlap with the pristine one, a sign of the difficulty to recombine all the created vacancies; this can be attributed to both intrinsic thermodynamical reasons \cite{Mickel2014} and to the difficulty of heating uniformly the whole defected structure by the electrical current, in contrast to what can be obtained by annealing treatment \cite{Xie2014a}. However, the restored metallic behavior is observed down to low temperature, as clearly reported in Fig.\,\ref{fig:Fig2}(b), where reversible switching between insulating and metallic states is achieved.
	
	\begin{figure*}[t]
		\begin{minipage}[c]{\textwidth}
			{
				\begin{minipage}[t]{0.62\textwidth}
					\mbox{}\\[-\baselineskip]
					\includegraphics[width=1.0\textwidth]{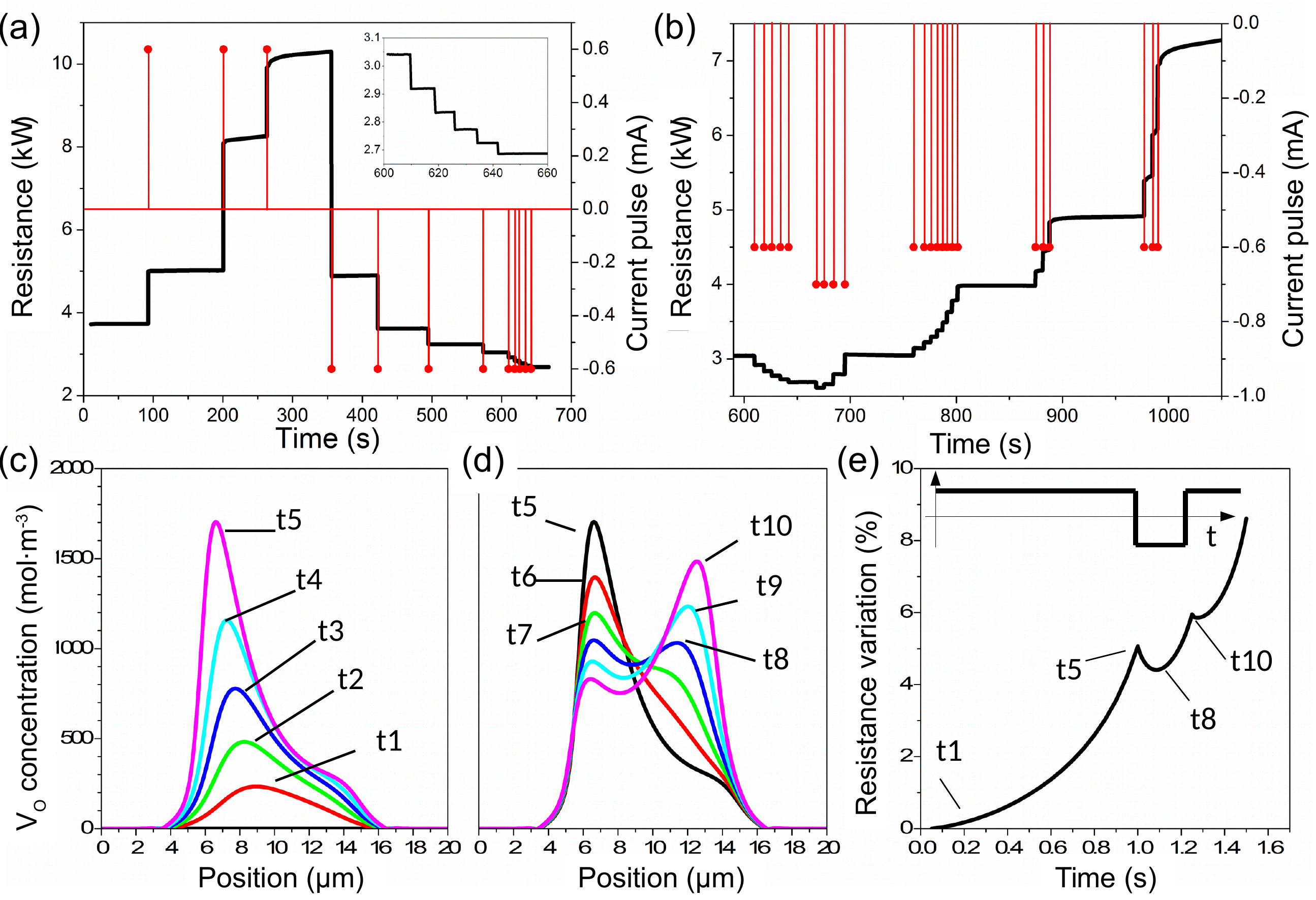}
				\end{minipage}\hfill
				\begin{minipage}[t]{0.33\textwidth}
					\mbox{}\\[-2\baselineskip]
					\caption{\label{fig:Fig3}
						Device electrical resistance under a sequence of current pulses of different polarity: experimental data ((a) and (b)) and proposed FEA model ((c)--(e)).
						(a) Pulses (drop-lines) of the same polarity determine an increase of resistance and if polarity is reversed the electrical resistance initially decreases (the inset is a magnification of the same plot).
						(b) More pulses with reversed polarization determine a new resistance increase. (c) Time-domain finite element analysis of oxygen vacancies profile evolution during an initial pulse of +370\,$\mu$A and (d) a reverse pulse of --370\,$\mu$A (total simulated time in (c) is 1\,s, step 200\,ms and in (d) is 0.25\,s, step\,50 ms). (e) Time evolution of the electrical resistance of the microbridge corresponding to (c) and (d), on top a schematic representation of the biasing sequence is presented.
					}
				\end{minipage}
			}
		\end{minipage}
	\end{figure*}
	
	Tuning the physical state of a perovskite oxide in a micrometric region through defect engineering by current pulses is intriguing, but it must face with the stability of the $\mathrm{V_O}$ profile against temperature-activated diffusion and electromigration phenomena. Indeed, oxidation/reduction mechanisms at oxide surfaces have been observed even at room temperature \cite{Buzio2012}. In our structure, we clearly observe slow drifts of the electrical resistance when the bridge is placed in	air or even in low vacuum. An increase of temperature and higher oxygen pressure condition determine faster evolution of the oxidation/diffusion processes (See Supplementary Material).
	
	The high mobility of oxygen vacancies in LSMO films	at temperatures slightly above ambient condition, together with the thermal characteristics of our microbridge allow to	uniquely exploit the coupling between current-induced heating and temperature-enhanced field-induced ion migration to achieve dynamical control of the physical state of an oxide on short time scales. The stability of the written state is related with the inhibition of $\mathrm{V_O}$ recombination and diffusion mechanisms; thus, it can be enhanced by vacuum condition and/or low temperatures. For example, R(T) plots of Fig.\,\ref{fig:Fig2}(b) remained stable through the whole time required for data acquisition (few hours).
	
	In the following, we describe how $\mathrm{V_O}$ profile along the microbridge can be modified by current pulses. Starting from
	a fully oxidized LSMO microbridge placed in vacuum environment ($\mathrm{2\cdot10^{-1}}$\,Pa) at 300\,K, we sent a ``forming'' current pulse (about 1\,mA for 1\,s, depending on geometry) to create oxygen vacancies with a consequent increase of resistance.
	We note that a second current pulse with opposite polarization and similar amplitude as the forming one lowers the measured resistance value. This decrease of resistance is observed for pulses of different durations, with shorter ones (<1\,s) determining smaller variations. If considering Joule-activated oxygen exchanges with the environment in vacuum conditions only, we should see a further increase of resistance. We interpret the observed behavior as a modification of the vacancies profile inside the microbridge, activated by combined Joule effect and electric field. To verify our model, we performed a series of measurements with current pulses	of different amplitude, while maintaining or inverting the pulse polarity with respect to the previous one (Fig.\,\ref{fig:Fig3}(a) and (b)). Our results clearly show that:
	\setlength{\leftmargini}{1.5 em}
	\begin{itemize}
		\setlength{\itemindent}{-0.5em}
		\item 	Starting from a pristine LSMO microbridge, the first
		increase of resistance is independent from the current polarity, because of the initial symmetry of the system.
		\item Following current pulses of the same amplitude and polarity of the initial one increase the measured resistance (Fig.\,\ref{fig:Fig3}(a)).
		\item When pulse polarity is inverted, the measured resistance is progressively lowered (Fig.\,\ref{fig:Fig3}(a) and (b)).
		\item If persisting with applying inverted pulses, the resistance starts increasing again (Fig.\,\ref{fig:Fig3}(b)).
	\end{itemize}

	This behavior is completely symmetric with respect to the initial pulse sign and only relative sign changes play a role. This phenomenology is observed in microbridges with different aspect ratio. The temperature profile along the microbridge affects the onset of oxygen exchange with the ambient and its diffusion through the thin film structure. At a given bias current, the temperature profile depends on microbridge dimensions and thermal dissipation with the environment (Ref. 25).
	
	In order to understand the main mechanisms of device operation, we developed a simple 2D FEA model (Comsol Multiphysics 4.3b) based on the linear approximation of $\mathrm{V_O}$	dynamics described in the review article of Merkle and	Maier \cite{Merkle2008}. We just considered a longitudinal section of a LSMO microbridge having 20\,$\mu$m of length and 100\,nm of thickness. Exchange with the environment is determined by the surface reaction rate ($\mathrm{k^\delta}$) and bulk electro-migration by the diffusion coefficient $D$. Temperature-dependence of both values are calculated by using an Arrhenius-type equations having energy barriers of 0.6\,eV and 1.2\,eV, respectively \cite{Yasuda1996,DeSouza2006}. For sake of simplicity, we considered the electrical conductivity as temperature-independent and related to the oxygen vacancies concentration by a second-order polynomial relationship. This is the most simple non-linear equation that could introduce a dependence between the overall resistance and the $\mathrm{V_O}$ profile shape. Further detail of our FEA model can be found in the Supplementary Material.
	
	Figure \ref{fig:Fig3}(c) shows the simulated time evolution of $\mathrm{V_O}$ distribution along the microbridge by an initial current pulse
	applied in vacuum conditions (typ. 370\,$\mu$A, 0.8\,mW, duration 1\,s). The calculated $\mathrm{V_O}$ profile evolves asymmetrically due to electromigration and the created oxygen vacancies are accumulated at one edge of the bridge, as shown, where
	the ion diffusion coefficient drops because of the lower temperature. Our device works as a sort of oxide solid state cell \cite{Baiatu1990}, with the role of the electrodes played by the microbridge edges, where oxygen vacancies are blocked. Instead, in the central region (which measures approximately one third of its total length) oxygen vacancies are continuously created and moved away, realizing an open system with	``doping storage'' capabilities. If the bias is inverted in sign, with roughly the same magnitude ($-370\,\mu$A), we observe a transfer of oxygen vacancies from one edge to the other	(Fig.\,\ref{fig:Fig3}(d)). The total amount of vacancies accumulating on the right part of the microbridge is given by both the $\mathrm{V_O}$	coming from the opposite side, which readily cross the high	diffusivity central region, and the newly created ones, coming from the high temperature surface at bridge center. The corresponding time evolution of the device electrical resistance is shown in Fig.\,\ref{fig:Fig3}(e) and qualitatively reproduces the observed behavior of Fig.\,\ref{fig:Fig3}(a) and (b), where, after the
	change of polarity, the resistance initially decreases before increasing again. We note how in Fig.\,\ref{fig:Fig3}(e) the resistance	evolution is calculated during the application of the current bias (bridge heated at high temperature), while in Fig.\,\ref{fig:Fig3}(a) and (b) the real device is probed after the current pulse with a small reading current (bridge nearby ambient temperature). In this last case, the resulting changes of resistance are amplified by the subsequent modification of the R \textit{vs} T characteristics of LSMO.
	
	Our basic model thus suggests that the observed drifts can be attributed to an electric-field driven redistribution of oxygen vacancies across the microbridge. The synergy between the electric field and the strong diffusivity gradients along the microbridge, determined by the inhomogeneous temperature profile, enables a peculiar feature of our system: the current pulse realizes $\mathrm{V_O}$ configurations that are far from equilibrium conditions and not accessible by the uniform heating procedures based on the use of furnaces or external heaters. Optimization of device operation would require the engineering of the best geometry in order to balance the efficiency of Joule heating with that of the electric field which drives $\mathrm{V_O}$ along the planar structure.
	
	Due to the small mass of the device, the fast drop of temperature after the current pulse rapidly suppresses oxygen dynamics, nearly freezing the created vacancies state with consequent result of increasing its lifetime. For longer times, the restoration of equilibrium condition is expected, but in vacuum environment and at ambient temperature the system is stable for several hours. The behavior of the system in terms of $\mathrm{V_O}$ distribution arises from the complex interplay between current magnitude, local conductivity, and temperature dependence of $D$ and $\mathrm{k^\delta}$. The values of their energy barriers suffer from significant experimental uncertainties and small deviations of their values can determine strong modifications of calculated profiles (see Supplementary Material). This, together with the previously mentioned strong simplifying assumption on conductivity modelling, makes our simulations just an attempt to give a phenomenological description of microbridge dynamics. Further investigations of the system would require the knowledge of its behavior at local scale, in particular, the possibility of formation of conductive filamentary	paths, which could be investigated in the future by local probe methods.
	
	In summary, we changed oxygen vacancies concentration in a free-standing manganite film by an electrical current applied in a controlled environment, reversibly switching the device from metallic to insulating behavior. The modified regions are confined within the microbridge length and can be switched with small power and short timescales. Our device works as a sort of oxygen vacancies accumulator, where creation and storage of defects are realized in different regions. Our results suggest that the electric field produced	by the current bias plays a major role in modifying the $\mathrm{V_O}$ profile shape and can be thus used to program multiple resistance states or even as an efficient tool to reconfigure the physical properties of many other oxide systems.
	\\
	\\
	We acknowledge financial support from MIUR through FIRB RBAP115AYN ``Oxides at the nanoscale: multifunctionality and applications'' and PRIN 2010NR4MXA ``OXIDE'' and from Universit\'a di Genova. We also acknowledge useful discussions with Emilio Bellingeri, Renato Buzio, Valentina Ceriale, Andrea Gerbi, Ilaria Pallecchi, Federico Remaggi and Antonio Sergio Siri.
	
	\bibliographystyle{apsrev4-1}
	\bibliography{library.bib}	

\end{document}